\documentclass[fleqn,twoside]{article}
\usepackage{espcrc2}

\usepackage{graphicx}
\usepackage[figuresright]{rotating}

\usepackage{scalefnt}

\newcommand{\AmS}{{\protect\the\textfont2
  A\kern-.1667em\lower.5ex\hbox{M}\kern-.125emS}}

\newcommand{\ep}{\varepsilon}

\newcommand{\be}{\begin{equation}}
\newcommand{\ee}{\end{equation}}
\newcommand{\bea}{\begin{eqnarray}}
\newcommand{\eea}{\end{eqnarray}}

\newcommand{\gm}{\gamma}

\newcommand{\dd}{\mbox{d}}

\newcommand{\nn}{\nonumber}

\title{The static quark potential to three loops in perturbation theory}

\author{Alexander V. Smirnov\address[1]{Scientific Research
    Computing Center of Moscow State University, Russia
  },
  Vladimir A. Smirnov\address[2]{Nuclear Physics Institute of Moscow
    State University, Russia
  },
  Matthias Steinhauser\address[3]{Institut f\"ur Theoretische
    Teilchenphysik, Karlsruhe Institute of Technology, Germany
  }\thanks{Talk presented at Loops and Legs in Quantum Field Theory 2010,
    W\"orlitz, Germany, April 25-30, 2010}
}

\begin{document}

\begin{abstract}
  The static potential constitutes a fundamental quantity of Quantum
  Chromodynamics. It has recently been evaluated to three-loop accuracy.
  In this contribution we provide details on the calculation and 
  present results for the 14 master integrals which contain a massless
  one-loop insertion.
\vspace{1pc}
\end{abstract}

% typeset front matter (including abstract)
\maketitle

%- {{{ Introduction:

\section{Introduction}

The static potential enters a variety of observables connected to heavy-quark
physics. Among them are the prominant examples like the determination of the
bottom quark mass from $\Upsilon$ sum rules or the cross section for the top
quark pair production close to theshold. It is desirable for both quantities
to perform a third-order analysis which requires the evaluation of the static
quark potential to three loops.

In order to fix the notation we write the static potential in momentum space
in the following form
\begin{eqnarray}
  \lefteqn{V(|{\vec q}\,|)=
  -{4\pi C_F\alpha_s(|{\vec q}\,|)\over{\vec q}\,^2}
  \Bigg[1+{\alpha_s(|{\vec q}\,|)\over 4\pi}a_1
  }
  \nonumber\\&&\mbox{}
  +\left({\alpha_s(|{\vec q}\,|)\over 4\pi}\right)^2a_2
  +\left({\alpha_s(|{\vec q}\,|)\over 4\pi}\right)^3
  \Bigg(a_3
  \nonumber\\&&\mbox{}
    + 8\pi^2 C_A^3\ln{\mu^2\over{\vec q}\,^2}\Bigg)
  +\cdots\Bigg]\,.
  \label{eq::V}
\end{eqnarray}
where $C_A=N_c$ and $C_F=(N_c^2-1)/(2N_c)$.
In Eq.~(\ref{eq::V}) we identify the renormalization scale $\mu^2$ and the
momentum transfer 
${\vec q}\,^2$. The complete dependence on $\mu$ can easily be restored with
the help of Eq.~(2) of Ref.~\cite{Smirnov:2008pn}.

The one- and two-loop coefficients 
$a_1$~\cite{Fischler:1977yf,Billoire:1979ih} 
and $a_2$~\cite{Peter:1996ig,Peter:1997me,Schroder:1998vy,Kniehl:2001ju} 
are given in Eq.~(4) of Ref.~\cite{Smirnov:2008pn} where also the
higher order terms in 
$\varepsilon$, necessary for the three-loop calculation, are presented.
In 2008 the fermionic contribution~\cite{Smirnov:2008pn} 
to $a_3$ was computed and end of 2009
$a_3$ was completed by evaluating also the purely gluonic part
which was 
achieved by two independent groups~\cite{Smirnov:2009fh,Anzai:2009tm}.
In this contribution we provide some
details of Ref.~\cite{Smirnov:2009fh} and in particular present explicit results for the
14 master integrals containing a massless one-loop sub-diagram.
The results for 16 more complicated integrals 
have been presented in Ref.~\cite{Smirnov:2010zc}.
In addition one has one more finite master integral which is only known
numerically and ten integrals which have no static line and are thus known
since long.

%- }}}
%- {{{ Reduction to MIs:

\section{Reduction to master integrals}

\begin{figure}[t]
  \begin{center}
    \begin{tabular}{ccc}
      \hspace*{-1em}
      \includegraphics[width=.16\textwidth]{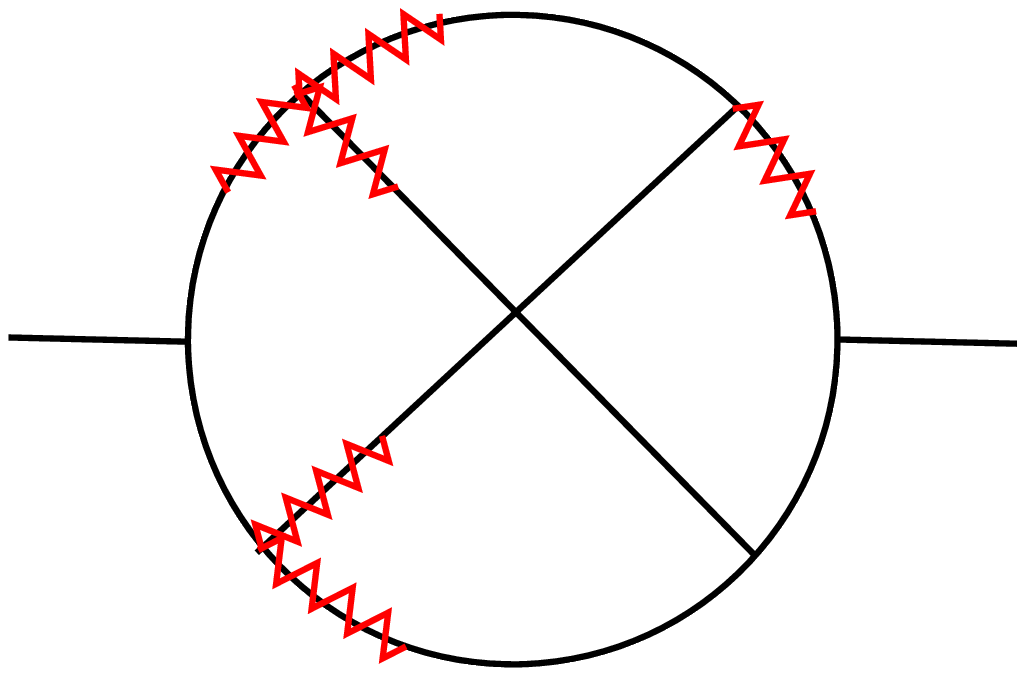} &
      \hspace*{-2em}
      \includegraphics[width=.16\textwidth]{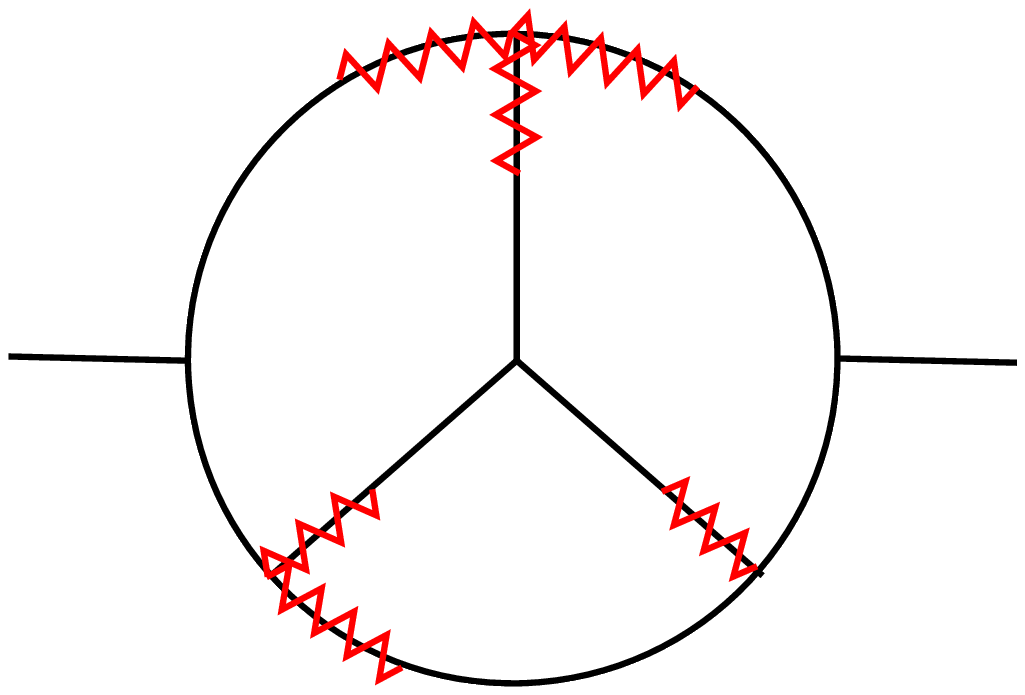} &
      \hspace*{-2em}
      \includegraphics[width=.16\textwidth]{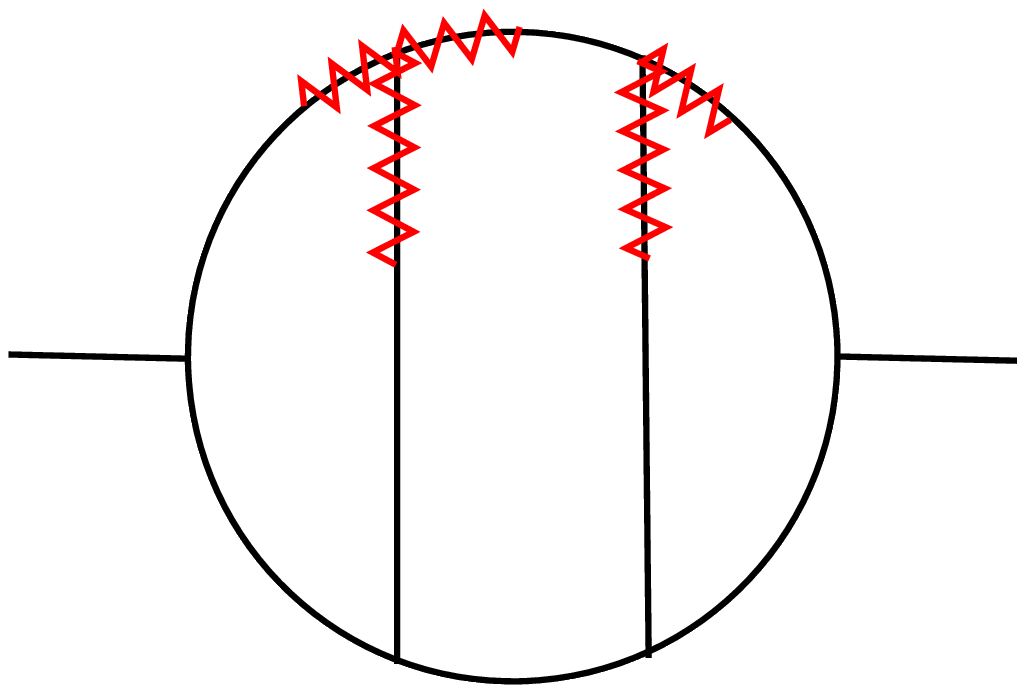}
    \end{tabular}

    \vspace*{-1em}

    \caption[]{\label{fig::j11abc}
      Three-loop two-point integrals with massless relativistic (solid lines) 
      and static propagators (zig-zag lines).
      }

    \vspace*{-2em}

 \end{center}
\end{figure}

In order to compute the static potential one has to consider the four-point
amplitudes describing the quark anti-quark interation. After integrating out
the heavy quark mass one arrives at non-relativistic QCD and remains with
only the
dependence on one kinematical variable, the momentum transfer between the
quarks. Consequently all occuring integrals can be mapped to one of the three
integrals displayed in Fig.~\ref{fig::j11abc}. We have performed both the direct
reduction of these integrals but also applied a partial fractioning in all
cases where three static lines meet in one vertex. This reduces the number of
indices which have to be considered during the reduction from 15 to twelve.

In Ref.~\cite{Smirnov:2009fh} the evaluation of $a_3$ has been performed for
general gauge parameter $\xi$ which is a very important check, in particular
for such involved calculations as the present one. However, the computational
price one has to pay is quite high: a rough estimate of the complexity
based on the number of integrals which have to be reduced to masters shows
that the linear $\xi$ term is about seven times and the $\xi^3$ term even
18 times more complicated than the Feynman gauge result.
Let us mention that nevertheless all occurring integrals could be reduced with
the help of {\tt FIRE}~\cite{FIRE}.

%- }}}
%- {{{ Results for master integrals:

\section{Results for master integrals}

\begin{figure}[t]
  \begin{center}
    \includegraphics[width=.45\textwidth]{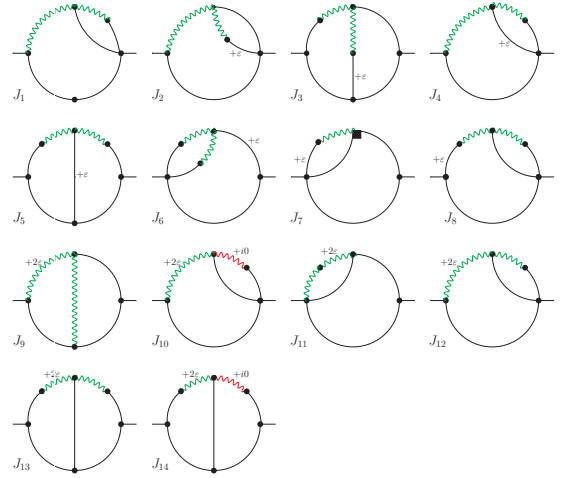}

    \vspace*{-1em}

   \caption[]{\label{fig::diags1}
     Two-loop master integrals with a regularized line
     contributing to $a_3$. The solid and zig-zag lines
     correspond to relativistic and static propagators, respectively.
     The black box in the case of $J_7$ denotes a
     monomial in the numerator.}
 \end{center}

    \vspace*{-2em}

\end{figure}

There are altogether 41 master integrals contributing to $a_3$. As already
mentioned above, in this contribution we want to consider those with a
massless one-loop insertion which can easily be integrated in terms of $\Gamma$
functions using standard
formulae. As a result one obtains the two-loop integrals in
Fig.~\ref{fig::diags1} where one of the indices has a non-integer exponent
involving the space-time
parameter $\varepsilon=(4-d)/2$ which is explicitly indicated next
to the corresponding line.
We present the analytical results for all integrals which can
be expressed as one of the following functions
%%%%%%%%%%%%%%%%%%%%%%%%%%%%%%%%%%%%%%%%
\bea
G^{(1)}_{a_1,\ldots,a_{7}}=\int\int\frac{\dd k\;\dd l}
{(-k^2)^{a_1}(-(k+q)^2)^{a_2}(-l^2)^{a_3}}
&& \nn \\ &&  \hspace*{-77mm}
\times
\frac{1}{(-(l+q)^2)^{a_4}(-(k-l)^2)^{a_5}
(-v\cdot k)^{a_6}(-v\cdot l)^{a_7}}\,,
\nn \\ &&  \hspace*{-77mm}
G^{(2)}_{a_1,\ldots,a_{7}}=\int\int\frac{\dd k\;\dd l}
{(-k^2)^{a_1}(-(k+q)^2)^{a_2}(-l^2)^{a_3}}
\nn \\ &&  \hspace*{-77mm}
\times
\frac{(-v\cdot (k-l))^{-a_7}}{(-(l+q)^2)^{a_4}(-(k-l)^2)^{a_5+\ep}
(-v\cdot k)^{a_6}}\,,
\nn \\ &&  \hspace*{-77mm}
G^{(3)}_{a_1,\ldots,a_{7}}=\int\int\frac{\dd k\;\dd l}
{(-k^2)^{a_1}(-(k+q)^2)^{a_2}(-l^2)^{a_3}}
\nn \\ &&  \hspace*{-77mm}
\times
\frac{(-v\cdot l)^{-a_7}}{(-(l+q)^2)^{a_4}(-(k-l)^2)^{a_5+\ep}
(-v\cdot k)^{a_6}}\,,
\nn \\ &&  \hspace*{-77mm}
G^{(4)}_{a_1,\ldots,a_{7}}=\int\int\frac{\dd k\;\dd l}
{(-k^2)^{a_1}(-(k+q)^2)^{a_2}(-l^2)^{a_3+\ep}}
\nn \\ &&  \hspace*{-77mm}
\times
\frac{(-v\cdot (k-l))^{-a_7}}{(-(l+q)^2)^{a_4}(-(k-l)^2)^{a_5}
(-v\cdot k)^{a_6}}\,,
\nn \\ &&  \hspace*{-77mm}
G^{(5)}_{a_1,\ldots,a_{7}}=\int\int\frac{\dd k\;\dd l}
{(-k^2)^{a_1+\ep}(-(k+q)^2)^{a_2}(-l^2)^{a_3}}
\nn \\ &&  \hspace*{-77mm}
\times
\frac{1}{(-(l+q)^2)^{a_4}(-(k-l)^2)^{a_5}
(-v\cdot k)^{a_6}(-v\cdot l)^{a_7}}\,,
\nn \\ &&  \hspace*{-77mm}
G^{(6)}_{a_1,\ldots,a_{7}}=\int\int\frac{\dd k\;\dd l}
{(-k^2)^{a_1}(-(k+q)^2)^{a_2}(-l^2)^{a_3}}
\nn \\ &&  \hspace*{-77mm}
\times
\frac{(-v\cdot (k-l))^{-a_7}}{(-(l+q)^2)^{a_4}(-(k-l)^2)^{a_5}
(-v\cdot k)^{a_6+2\ep}}\,,
\nn \\ &&  \hspace*{-77mm}
G^{(7,\pm)}_{a_1,\ldots,a_{7}}=\int\int\frac{\dd k\;\dd l}
{(-k^2)^{a_1}(-(k+q)^2)^{a_2}(-l^2)^{a_3}}
\nn \\ &&  \hspace*{-77mm}
\times
\frac{(-v\cdot l\mp i0)^{-a_7}}{(-(l+q)^2)^{a_4}(-(k-l)^2)^{a_5}
(-v\cdot k)^{a_6+2\ep}}
\,,\nn
\eea
%%%%%%%%%%%%%%%%%%%%%%%%%%%%%%%%%%%%%%%%
where in all propagators, apart the last one, the causal $-i0$ is implied.
 
In the following we set for simplicity $q^2=-1$ and $v^2=1$ since the
corresponding dependences can easily be reconstructed.
The results for $J_1$ to $J_{14}$ read
%%%%%%%%%%%%%%%%%%%%%%%%%%%%%%%%%%%%%%%%
\bea
%G(420,\{0,1,1,0,1,1,1\})=
J_1=G^{(1)}_{0,1,1,0,1,1,1}=
\pi^2
\Bigg[
\frac{2}{3 \ep}+4+\left(24-\frac{7 \pi ^2}{9}\right)\ep
&& \nn \\ &&  \hspace*{-79mm}
+\left(144-\frac{14 \pi^2}{3}-\frac{352 \zeta (3)}{9}\right) \ep^2
+\left(864
\right.
\nn \\ &&  \hspace*{-79mm}
\left.
-28 \pi^2-\frac{101 \pi ^4}{60}-\frac{704 \zeta (3)}{3}\right) \ep^3
 +\ldots\Bigg],
\nn
\eea
%G(430,\{0,1,1,0,1,1,1\})=
\bea
J_2=G^{(2)}_{0,1,1,0,1,1,1}=
\frac{4 \pi ^2}{9\ep}
+\frac{4 \zeta (3)}{3}+\frac{32 \pi ^2}{9}
&& \nn \\ &&  \hspace*{-70mm}
+\left(\frac{256 \pi ^2}{9}-\frac{34 \pi ^4}{45}+\frac{32 \zeta
   (3)}{3}\right) \ep
+\left(\frac{2048 \pi ^2}{9}
\right.
\nn \\ &&  \hspace*{-70mm}
\left.
-\frac{272\pi ^4}{45} +\frac{256 \zeta (3)}{3}-\frac{1622 \pi ^2 \zeta (3)}{27}
-\frac{20 \zeta(5)}{3}\right) \ep^2
\nn \\ &&  \hspace*{-70mm}
+\left(\frac{16384 \pi ^2}{9}-\frac{2176 \pi ^4}{45}-\frac{2339\pi ^6}{630}
\right. \nn \\ &&  \hspace*{-70mm}
+\frac{2048 \zeta (3)}{3}-\frac{12976 \pi ^2 \zeta (3)}{27}-\frac{968 \zeta(3)^2}{9}
\nn \\ &&  \hspace*{-70mm}
\left.
-\frac{160 \zeta (5)}{3}\right) \ep^3
+\ldots,
\nn
\eea
\bea
%G(430,\{1,1,1,1,1,1,1\})=
J_3=G^{(2)}_{1,1,1,1,1,1,1}=
-\frac{8 \pi ^2}{9 \ep}-\frac{20 \zeta (3)}{3}-\frac{16 \pi^2}{9}
&& \nn \\ &&  \hspace*{-74mm}
+\left(\frac{256 \pi ^2}{9}-\frac{106 \pi^4}{45}-\frac{112 \zeta (3)}{3}\right) \ep
   \nn \\ &&  \hspace*{-74mm}
+\left(-\frac{1792 \pi ^2}{9}+\frac{688 \pi
   ^4}{45}-18 \pi ^4 \log (2)+\frac{1216 \zeta (3)}{3}
\right.   \nn \\ &&  \hspace*{-74mm}
\left.
   +\frac{5530 \pi ^2 \zeta
   (3)}{27}-\frac{1865 \zeta (5)}{3}\right) \ep^2
\nn \\ &&  \hspace*{-74mm}
+\left(-864 s_6+\frac{2660 \zeta (5)}{3}
+\frac{12778 \zeta (3)^2}{9}
\right.
\nn \\ &&  \hspace*{-74mm}
\left.
-\frac{2440 \pi ^2 \zeta
   (3)}{27}-\frac{7936 \zeta (3)}{3}-576 \pi ^2 \mbox{Li}_4\left(\frac{1}{2}\right)
\right.  \nn \\ &&  \hspace*{-74mm}
   -24 \pi ^2
   \log ^4(2)-30 \pi ^4 \log ^2(2)+72 \pi ^4 \log (2)
\nn \\ &&  \hspace*{-74mm}
\left.
   +\frac{392 \pi ^6}{135}-\frac{3808 \pi
   ^4}{45}+\frac{10240 \pi ^2}{9}\right) \ep^3
+\ldots,
\nn
\eea
%G(440,\{0,1,1,0,1,1,1\})=
\bea
J_4=G^{(3)}_{0,1,1,0,1,1,1}=
\frac{4 \pi ^2}{9\ep}-\frac{8 \zeta (3)}{3}+\frac{32 \pi ^2}{9}
&& \nn \\ &&  \hspace*{-70mm}
 +\left(\frac{256 \pi ^2}{9}-\frac{14 \pi ^4}{15}-\frac{64 \zeta
   (3)}{3}\right) \ep
+\left(\frac{2048 \pi ^2}{9}
\right.
  \nn \\ &&  \hspace*{-70mm}
\left.
-\frac{112\pi ^4}{15}   -\frac{512 \zeta (3)}{3}-\frac{1604 \pi ^2 \zeta (3)}{27}+\frac{40 \zeta
   (5)}{3}\right) \ep^2
  \nn \\ &&  \hspace*{-70mm}
+\left(\frac{16384 \pi ^2}{9}-\frac{896 \pi ^4}{15}-\frac{6161
   \pi ^6}{1890}-\frac{4096 \zeta (3)}{3}
\right.
\nn \\ &&  \hspace*{-70mm}
\left.
   -\frac{12832 \pi ^2 \zeta (3)}{27}+\frac{1936 \zeta
   (3)^2}{9}+\frac{320 \zeta (5)}{3}\right) \ep^3
+\ldots,
\nn
\eea
\bea
%G(440,\{1,1,1,1,1,1,1\})=
J_5=G^{(3)}_{1,1,1,1,1,1,1}=
-\frac{8 \pi ^2}{9 \ep}+\frac{40 \zeta (3)}{3}-\frac{16 \pi^2}{9}
&& \nn \\ &&  \hspace*{-74mm}
+\left(\frac{256 \pi ^2}{9}-\frac{12 \pi
   ^4}{5}+\frac{224 \zeta (3)}{3}\right) \ep
+\left(-\frac{1792 \pi ^2}{9}
\right.
\nn \\ &&  \hspace*{-74mm}
+\frac{136 \pi^4}{5}-36 \pi ^4 \log (2)-\frac{2432 \zeta (3)}{3}
+\frac{5440 \pi ^2 \zeta(3)}{27}
\nn \\ &&  \hspace*{-74mm}
\left.
   +\frac{3730 \zeta (5)}{3}\right) \ep^2
+\left(1728
   s_6-\frac{5320 \zeta (5)}{3}
\right.
\nn \\ &&  \hspace*{-74mm}
   -\frac{25556 \zeta (3)^2}{9}-\frac{2944 \pi ^2 \zeta(3)}{27}
   +\frac{15872 \zeta (3)}{3}
\nn \\ &&  \hspace*{-74mm}
-576 \pi ^2 \mbox{Li}_4\left(\frac{1}{2}\right)
-24 \pi ^2\log ^4(2)-84 \pi ^4 \log ^2(2)
\nn \\ &&  \hspace*{-74mm}
\left.+144 \pi ^4 \log (2)
+\frac{188 \pi ^6}{27}-\frac{896 \pi^4}{5}+\frac{10240 \pi ^2}{9}\right) \ep^3
\nn \\ &&  \hspace*{-74mm}
+\ldots,
\nn
\eea
%G(450,\{1,0,1,1,1,1,1\})=
\bea
J_6=G^{(4)}_{1,0,1,1,1,1,1}=
-\frac{4 \pi ^2}{9\ep}-\frac{40 \zeta (3)}{3}+\frac{16 \pi ^2}{9}
&& \nn \\ &&  \hspace*{-74mm}
+\left(-\frac{64 \pi ^2}{9}+\frac{2 \pi ^4}{9}+\frac{160 \zeta
   (3)}{3}\right) \ep
 \nn \\ &&  \hspace*{-74mm}
+\left(\frac{256 \pi ^2}{9}-\frac{8 \pi
   ^4}{9}-\frac{640 \zeta (3)}{3}+\frac{1604 \pi ^2 \zeta (3)}{27}
\right.
   \nn \\ &&  \hspace*{-74mm}
\left.
-\frac{1120 \zeta(5)}{3}\right) \ep^2
+\left(-\frac{1024 \pi ^2}{9}+\frac{32 \pi ^4}{9}+\frac{3061 \pi^6}{1890}
\right.
 \nn \\ &&  \hspace*{-74mm}
  +\frac{2560 \zeta (3)}{3}-\frac{6416 \pi ^2 \zeta (3)}{27}+\frac{1400 \zeta
   (3)^2}{9}
  \nn \\ &&  \hspace*{-74mm}
\left.   +\frac{4480 \zeta (5)}{3}\right) \ep^3
+\ldots,
\nn
\eea
\bea
%G(460,\{1,0,1,1,1,1,-1\})=
J_7=G^{(5)}_{1,0,1,1,1,1,-1}=
\frac{1}{3 \ep}-2 \zeta (3)+4
&& \nn \\ &&  \hspace*{-62mm}
+\left(\frac{100}{3}-\frac{\pi ^2}{18}-\frac{\pi ^4}{10}-12 \zeta (3)\right)\ep
+\left(240-\frac{2 \pi ^2}{3}
\right.
\nn \\ &&  \hspace*{-62mm}
\left.
-\frac{3 \pi ^4}{5}-\frac{728 \zeta
   (3)}{9}+\frac{\pi ^2 \zeta (3)}{3}-52 \zeta (5)\right)
   \ep^2
\nn \\ &&  \hspace*{-62mm}
+\left(\frac{4816}{3}-\frac{50 \pi ^2}{9}-\frac{449 \pi ^4}{120}-\frac{319 \pi
   ^6}{1260}-\frac{1616 \zeta (3)}{3}
\right.
\nn \\ &&  \hspace*{-62mm}
\left.
   +2 \pi ^2 \zeta (3)+\frac{232 \zeta (3)^2}{3}-312 \zeta
   (5)\right) \ep^3
+\ldots,
\nn
\eea
%G(460,\{1,1,1,0,1,1,1\})=
\bea
J_8=G^{(5)}_{1,1,1,0,1,1,1}=
-\frac{4 \pi ^2}{9\ep}+\frac{20 \zeta (3)}{3}+\frac{16 \pi ^2}{9}
&& \nn \\ &&  \hspace*{-74mm}
+\left(-\frac{64 \pi ^2}{9}+\frac{16 \pi ^4}{9}-\frac{80 \zeta
   (3)}{3}\right) \ep
+\left(\frac{256 \pi ^2}{9}
\right.
  \nn \\ &&  \hspace*{-74mm}
\left.
 -\frac{64 \pi^4}{9}
   +\frac{320 \zeta (3)}{3}+\frac{1514 \pi ^2 \zeta (3)}{27}+\frac{560 \zeta
   (5)}{3}\right) \ep^2
 \nn \\ &&  \hspace*{-74mm}
+\left(-\frac{1024 \pi ^2}{9}+\frac{256 \pi ^4}{9}+\frac{6971 \pi
   ^6}{1890}-\frac{1280 \zeta (3)}{3}
\right.
\nn \\ &&  \hspace*{-74mm}
\left.
   -\frac{6056 \pi ^2 \zeta (3)}{27}-\frac{700 \zeta
   (3)^2}{9}-\frac{2240 \zeta (5)}{3}\right) \ep^3
+\ldots,
\nn
\eea
\bea
%G(475,\{0,1,1,1,0,1,1\})=
J_9=G^{(6)}_{0,1,1,1,0,1,1}=
\frac{1}{3\ep^2}+\left(\frac{8}{3}+\frac{2}{3} \log (2)\right)\frac{1}{\ep}
&&\nn \\ &&  \hspace*{-74mm}
+\frac{2 \log ^2(2)}{3}+\frac{16 \log (2)}{3}+\frac{2 \pi^2}{9}+\frac{52}{3}
\nn \\ &&  \hspace*{-74mm}
+\left(\frac{320}{3}+\frac{16 \pi ^2}{9}+\frac{104 \log (2)}{3}+\frac{4}{9} \pi ^2
   \log (2)
   \right.
\nn \\ &&  \hspace*{-74mm}
\left.
   +\frac{16 \log ^2(2)}{3}+\frac{4 \log ^3(2)}{9}-\frac{140 \zeta (3)}{9}\right)
   \ep
 \nn \\ &&  \hspace*{-74mm}
+\left(\frac{1936}{3}+\frac{104 \pi
   ^2}{9}-\frac{179 \pi ^4}{270}+\frac{640 \log (2)}{3}
\right.
\nn \\ &&  \hspace*{-74mm}
   +\frac{32}{9} \pi ^2 \log (2)+\frac{104
   \log ^2(2)}{3}+\frac{4}{9} \pi ^2 \log ^2(2)
  \nn \\ &&  \hspace*{-74mm}
 +\frac{32 \log ^3(2)}{9}  +\frac{2 \log^4(2)}{9}-\frac{1120 \zeta (3)}{9}
  \nn \\ &&  \hspace*{-74mm}
\left.
-\frac{280 \log (2) \zeta (3)}{9}\right)
   \ep^2
+\left(\frac{11648}{3}+\frac{640 \pi ^2}{9}
\right.
 \nn \\ &&  \hspace*{-74mm}
-\frac{716 \pi ^4}{135}+\frac{3872 \log
   (2)}{3}+\frac{208}{9} \pi ^2 \log (2)
\nn \\ &&  \hspace*{-74mm}
-\frac{179}{135} \pi ^4 \log (2)   +\frac{640 \log^2(2)}{3}
+\frac{32}{9} \pi ^2 \log ^2(2)
\nn \\ &&  \hspace*{-74mm}
 +\frac{208 \log ^3(2)}{9}
      +\frac{8}{27} \pi ^2 \log^3(2)+\frac{16 \log ^4(2)}{9}
   \nn \\ &&  \hspace*{-74mm}
      +\frac{4 \log ^5(2)}{45}-\frac{7280 \zeta (3)}{9}
   -\frac{280 \pi^2 \zeta (3)}{27}
  \nn \\ &&  \hspace*{-74mm}
   -\frac{2240 \log (2) \zeta (3)}{9}-\frac{280}{9} \log ^2(2) \zeta(3)
  \nn \\ &&  \hspace*{-74mm}
  \left.  -\frac{6572 \zeta (5)}{15}\right) \ep^3
+\ldots,
\nn
\eea
%G(478,\{0,1,1,0,1,1,1\})=
\bea
J_{10}=G^{(7,-)}_{0,1,1,0,1,1,1}=
-\frac{8 \pi ^2}{9 \ep}
-\frac{8 \zeta (3)}{3}
&& \nn \\ &&  \hspace*{-61mm}
-\frac{16}{9} \pi ^2 \log (2)-\frac{64 \pi^2}{9}
+\left(-\frac{512 \pi ^2}{9}+\frac{40 \pi ^4}{27}
\right.
\nn \\ &&  \hspace*{-61mm}
-\frac{128}{9} \pi ^2
   \log (2)-\frac{16}{9} \pi ^2 \log ^2(2)-\frac{64 \zeta (3)}{3}
  \nn \\ &&  \hspace*{-61mm}
\left.   -\frac{16 \log (2) \zeta
   (3)}{3}\right) \ep
 +\left(-\frac{4096 \pi ^2}{9}+\frac{320 \pi^4}{27}
\right.
\nn \\ &&  \hspace*{-61mm}
   -\frac{1024}{9} \pi ^2 \log (2)+\frac{80}{27} \pi ^4 \log (2)-\frac{128}{9} \pi ^2
   \log ^2(2)
 \nn \\ &&  \hspace*{-61mm}
   -\frac{32}{27} \pi ^2 \log ^3(2)
   -\frac{512 \zeta (3)}{3}
+\frac{3184 \pi ^2 \zeta(3)}{27}
\nn \\ &&  \hspace*{-61mm}
\left.
   -\frac{128 \log (2) \zeta (3)}{3}-\frac{16}{3} \log ^2(2) \zeta (3)+24 \zeta
   (5)\right) \ep^2
  \nn \\ &&  \hspace*{-61mm}
+\left(-\frac{32768 \pi ^2}{9}+\frac{2560 \pi ^4}{27}+\frac{52 \pi^6}{7}
\right.
  \nn \\ &&  \hspace*{-61mm}
\left.
   -\frac{8192}{9} \pi ^2 \log (2)+\frac{640}{27} \pi ^4 \log (2)-\frac{1024}{9} \pi ^2
   \log ^2(2)
  \right.  \nn \\ &&  \hspace*{-61mm}
 +\frac{80}{27} \pi ^4 \log ^2(2)
   -\frac{256}{27} \pi ^2 \log ^3(2)-\frac{16}{27} \pi
   ^2 \log ^4(2)
\nn \\ &&  \hspace*{-61mm}
   -\frac{4096 \zeta (3)}{3}+\frac{25472 \pi ^2 \zeta (3)}{27}-\frac{1024 \log (2)\zeta (3)}{3}
 \nn \\ &&  \hspace*{-61mm}
   +\frac{6368}{27} \pi ^2 \log (2) \zeta (3)
   -\frac{128}{3} \log ^2(2) \zeta(3)
 \nn \\ &&  \hspace*{-61mm}
   -\frac{32}{9} \log ^3(2) \zeta (3)+\frac{2032 \zeta (3)^2}{9}+192 \zeta (5)
 \nn \\ &&  \hspace*{-61mm}
 \left.
+48 \log (2)\zeta (5)\right) \ep^3
+\ldots,
\nn
\eea
%G(480,\{0,0,1,1,1,2,0\})=
\bea
J_{11}=G^{(7,+)}_{0,0,1,1,1,2,0}=
\frac{1}{3\ep^2}+\left(\frac{4}{3}+\frac{2 \log (2)}{3}\right)\frac{1}{\ep}
&&\nn \\ &&  \hspace*{-77mm}
+\frac{2 \log ^2(2)}{3}+\frac{8 \log (2)}{3}-\frac{2 \pi^2}{9}+\frac{28}{3}
+\left(\frac{160}{3}-\frac{8 \pi ^2}{9}
\right.
\nn \\ &&  \hspace*{-77mm}
+\frac{56 \log (2)}{3}-\frac{4}{9} \pi ^2
   \log (2)+\frac{8 \log ^2(2)}{3}+\frac{4 \log ^3(2)}{9}
  \nn \\ &&  \hspace*{-77mm}
\left.
   -\frac{188 \zeta (3)}{9}\right)\ep
+\left(\frac{976}{3}-\frac{56 \pi
   ^2}{9}-\frac{9 \pi ^4}{10}+\frac{320 \log (2)}{3}
 \right.
  \nn \\ &&  \hspace*{-77mm}
   -\frac{16}{9} \pi ^2 \log (2)+\frac{56 \log
   ^2(2)}{3}-\frac{4}{9} \pi ^2 \log ^2(2)
 \nn \\ &&  \hspace*{-77mm}
   +\frac{16 \log ^3(2)}{9}  +\frac{2 \log
   ^4(2)}{9}-\frac{752 \zeta (3)}{9}
 \nn \\ &&  \hspace*{-77mm}
 \left.   -\frac{376 \log (2) \zeta (3)}{9}\right)
   \ep^2
+\left(\frac{5824}{3}-\frac{320 \pi ^2}{9}-\frac{18 \pi ^4}{5}
\right.
\nn \\ &&  \hspace*{-77mm}
 +\frac{1952 \log(2)}{3}  -\frac{112}{9} \pi ^2 \log (2)-\frac{9}{5} \pi ^4 \log (2)
\nn \\ &&  \hspace*{-77mm}
+\frac{320 \log^2(2)}{3}
 -\frac{16}{9} \pi ^2 \log ^2(2)
   +\frac{112 \log ^3(2)}{9}
 \nn \\ &&  \hspace*{-77mm}
   -\frac{8}{27} \pi ^2 \log
   ^3(2)+\frac{8 \log ^4(2)}{9}+\frac{4 \log ^5(2)}{45}
  \nn \\ &&  \hspace*{-77mm} \left.
   -\frac{5264 \zeta (3)}{9}+\frac{376 \pi
   ^2 \zeta (3)}{27}-\frac{1504 \log (2) \zeta (3)}{9}
 \right.  \nn \\ &&  \hspace*{-77mm} \left.
     -\frac{376}{9} \log ^2(2) \zeta(3)-\frac{7532 \zeta (5)}{15}\right) \ep^3
+\left(\frac{35008}{3}
\right.
 \nn \\ &&  \hspace*{-77mm}
-\frac{1952 \pi ^2}{9}-\frac{126 \pi ^4}{5}-\frac{5617 \pi
   ^6}{2835}+\frac{11648 \log (2)}{3}
  \nn \\ &&  \hspace*{-77mm}
   -\frac{640}{9} \pi ^2 \log (2)-\frac{36}{5} \pi ^4 \log
   (2)+\frac{1952 \log ^2(2)}{3}
 \nn \\ &&  \hspace*{-77mm}
   -\frac{112}{9} \pi ^2 \log ^2(2)-\frac{9}{5} \pi ^4 \log
   ^2(2)+\frac{640 \log ^3(2)}{9}
  \nn \\ &&  \hspace*{-77mm} \left.
   -\frac{32}{27} \pi ^2 \log ^3(2)+\frac{56 \log
   ^4(2)}{9}-\frac{4}{27} \pi ^2 \log ^4(2)
  \right.  \nn \\ &&  \hspace*{-77mm}
+\frac{16 \log ^5(2)}{45}
+\frac{4 \log^6(2)}{135}-\frac{30080 \zeta (3)}{9}
\nn \\ &&  \hspace*{-77mm}
+\frac{1504 \pi ^2 \zeta (3)}{27}-\frac{10528 \log (2)\zeta (3)}{9}
\nn \\ &&  \hspace*{-77mm} \left.
   +\frac{752}{27} \pi ^2 \log (2) \zeta (3)-\frac{1504}{9} \log ^2(2) \zeta(3)
 \right.  \nn \\ &&  \hspace*{-77mm}
   -\frac{752}{27} \log ^3(2) \zeta (3)+\frac{17672 \zeta (3)^2}{27}-\frac{30128 \zeta
   (5)}{15}
  \nn \\ &&  \hspace*{-77mm} \left.
   -\frac{15064 \log (2) \zeta (5)}{15}\right)\ep^4
+\ldots,
\nn
\eea
%G(480,\{0,1,1,0,1,1,1\})=
\bea
J_{12}=G^{(7,+)}_{0,1,1,0,1,1,1}=
+\frac{4 \pi ^2}{9 \ep}
-\frac{8 \zeta (3)}{3}+\frac{8}{9} \pi ^2 \log (2)
\nn \\ &&  \hspace*{-80mm}
 +\frac{32 \pi^2}{9}  +\left(\frac{256 \pi ^2}{9}-\frac{20 \pi ^4}{27}+\frac{64}{9} \pi ^2 \log(2)
 \right.
   \nn \\ &&  \hspace*{-80mm}
\left.
   +\frac{8}{9} \pi ^2 \log ^2(2)-\frac{64 \zeta (3)}{3}-\frac{16 \log (2) \zeta
   (3)}{3}\right) \ep
  \nn \\ &&  \hspace*{-80mm}
+\left(\frac{2048 \pi ^2}{9}-\frac{160 \pi ^4}{27}+\frac{512}{9}
   \pi ^2 \log (2)
   \right.
 \nn \\ &&  \hspace*{-80mm}
\left.
   -\frac{40}{27} \pi ^4 \log (2)+\frac{64}{9} \pi ^2 \log ^2(2)+\frac{16}{27}
   \pi ^2 \log ^3(2)
   \right.  \nn \\ &&  \hspace*{-80mm}
-\frac{512 \zeta (3)}{3}
   -\frac{1520 \pi ^2 \zeta (3)}{27}-\frac{128 \log (2)
   \zeta (3)}{3}
   \nn \\ &&  \hspace*{-80mm} \left.
   -\frac{16}{3} \log ^2(2) \zeta (3)+24 \zeta (5)\right)
   \ep^2
+\left(\frac{16384 \pi ^2}{9}
\right.
 \nn \\ &&  \hspace*{-80mm}
-\frac{1280 \pi ^4}{27}-\frac{218 \pi
   ^6}{63}+\frac{4096}{9} \pi ^2 \log (2)
\nn \\ &&  \hspace*{-80mm}
\left.
   -\frac{320}{27} \pi ^4 \log (2)+\frac{512}{9} \pi ^2
   \log ^2(2)-\frac{40}{27} \pi ^4 \log ^2(2)
  \right.  \nn \\ &&  \hspace*{-80mm}
   +\frac{128}{27} \pi ^2 \log ^3(2)+\frac{8}{27} \pi
   ^2 \log ^4(2)-\frac{4096 \zeta (3)}{3}
  \nn \\ &&  \hspace*{-80mm}
   -\frac{12160 \pi ^2 \zeta (3)}{27}-\frac{1024 \log (2)
   \zeta (3)}{3}
  \nn \\ &&  \hspace*{-80mm}
    -\frac{3040}{27} \pi ^2 \log (2) \zeta (3)
-\frac{128}{3} \log ^2(2) \zeta(3)
   \nn \\ &&  \hspace*{-80mm}
   -\frac{32}{9} \log ^3(2) \zeta (3)+\frac{2032 \zeta (3)^2}{9}+192 \zeta (5)
   \nn \\ &&  \hspace*{-80mm}
 \left.
   +48 \log (2)
   \zeta (5)\right) \ep^3
+\ldots,
\nn
\eea
%G(480,\{1,1,1,1,1,1,1\})=
\bea
J_{13}=G^{(7,+)}_{1,1,1,1,1,1,1}=
-\frac{8 \pi ^2}{9 \ep}
+\frac{40 \zeta (3)}{3}
&&\nn \\ &&  \hspace*{-63mm}
-\frac{16}{9}\pi ^2 \log (2)-\frac{16 \pi ^2}{9}
+\left(\frac{256 \pi ^2}{9}-\frac{292 \pi^4}{135}
\right.
 \nn \\ &&  \hspace*{-63mm}
   -\frac{32}{9} \pi ^2 \log (2)-\frac{16}{9} \pi ^2 \log ^2(2)+\frac{224 \zeta
   (3)}{3}
  \nn \\ &&  \hspace*{-63mm}
\left.
   +\frac{80 \log (2) \zeta (3)}{3}\right) \ep
+\left(-\frac{1792 \pi
   ^2}{9}+\frac{3904 \pi ^4}{135}
\right.
  \nn \\ &&  \hspace*{-63mm}
\left.
   +\frac{512}{9} \pi ^2 \log (2)-\frac{4904}{135} \pi ^4 \log
   (2)-\frac{32}{9} \pi ^2 \log ^2(2)
 \right.  \nn \\ &&  \hspace*{-63mm}
  -\frac{32}{27} \pi ^2 \log ^3(2)
   -\frac{2432 \zeta
   (3)}{3}+\frac{4912 \pi ^2 \zeta (3)}{27}
\nn \\ &&  \hspace*{-63mm} \left.
   +\frac{448 \log (2) \zeta (3)}{3}+\frac{80}{3} \log
   ^2(2) \zeta (3)+1176 \zeta (5)\right) \ep^2
 \nn \\ &&  \hspace*{-63mm}
+\left(\frac{10240 \pi ^2}{9}-\frac{5312 \pi ^4}{27}+\frac{11908 \pi
   ^6}{2835}
\right.
   \nn \\ &&  \hspace*{-63mm}
\left.
   -\frac{3584}{9} \pi ^2 \log (2)+\frac{25088}{135} \pi ^4 \log (2)
  \right.  \nn \\ &&  \hspace*{-63mm}
   +\frac{512}{9} \pi
   ^2 \log ^2(2)-\frac{17864}{135} \pi ^4 \log ^2(2)
 \nn \\ &&  \hspace*{-63mm}
   -\frac{64}{27} \pi ^2 \log
   ^3(2)-\frac{16}{27} \pi ^2 \log ^4(2)
    +\frac{15872 \zeta (3)}{3}
  \nn \\ &&  \hspace*{-63mm} \left.
  +\frac{896 \pi ^2 \zeta
   (3)}{27}-\frac{4864 \log (2) \zeta (3)}{3}
 \right.  \nn \\ &&  \hspace*{-63mm}
+\frac{9824}{27} \pi ^2 \log (2) \zeta(3)
   +\frac{448}{3} \log ^2(2) \zeta (3)
   \nn \\ &&  \hspace*{-63mm}
   +\frac{160}{9} \log ^3(2) \zeta (3)-\frac{16496 \zeta
   (3)^2}{9}-864 \zeta (5)
  \nn \\ &&  \hspace*{-63mm} \left.
   +2352 \log (2) \zeta (5)\right) \ep^3
+\ldots,
\nn
\eea
%G(478,\{1,1,1,1,1,1,1\})=
\bea
J_{14}=G^{(7,-)}_{1,1,1,1,1,1,1}=
\frac{16 \pi ^2}{9 \ep}
+\frac{40 \zeta (3)}{3}
&&\nn \\ &&  \hspace*{-62mm}
+\frac{32}{9}\pi ^2 \log (2)+\frac{32 \pi ^2}{9}
+\left(-\frac{512 \pi ^2}{9}+\frac{548 \pi
   ^4}{135}
 \right.
  \nn \\ &&  \hspace*{-62mm}
   +\frac{64}{9} \pi ^2 \log (2)+\frac{32}{9} \pi ^2 \log ^2(2)+\frac{224 \zeta
   (3)}{3}
 \nn \\ &&  \hspace*{-62mm}
\left.
   +\frac{80 \log (2) \zeta (3)}{3}\right) \ep
+\left(\frac{3584 \pi
   ^2}{9}-\frac{4496 \pi ^4}{135}
\right.
   \nn \\ &&  \hspace*{-62mm}
 \left.
   -\frac{1024}{9} \pi ^2 \log (2)+\frac{5416}{135} \pi ^4 \log
   (2)+\frac{64}{9} \pi ^2 \log ^2(2)
 \right.  \nn \\ &&  \hspace*{-62mm}
+\frac{64}{27} \pi ^2 \log ^3(2)
   -\frac{2432 \zeta
   (3)}{3}-\frac{10544 \pi ^2 \zeta (3)}{27}
  \nn \\ &&  \hspace*{-62mm} \left.
   +\frac{448 \log (2) \zeta (3)}{3}+\frac{80}{3} \log
   ^2(2) \zeta (3)+1176 \zeta (5)\right) \ep^2
\nn \\ &&  \hspace*{-62mm}
+\left(-\frac{20480 \pi ^2}{9}+\frac{5440 \pi ^4}{27}-\frac{4472 \pi
   ^6}{2835}
\right.
\nn \\ &&  \hspace*{-62mm}
   +\frac{7168}{9} \pi ^2 \log (2)-\frac{26272}{135} \pi ^4 \log (2)
  \nn \\ &&  \hspace*{-62mm}
\left.
   -\frac{1024}{9} \pi
   ^2 \log ^2(2) +\frac{18376}{135} \pi ^4 \log ^2(2)
\right.  \nn \\ &&  \hspace*{-62mm}
  +\frac{128}{27} \pi ^2 \log
   ^3(2)+\frac{32}{27} \pi ^2 \log ^4(2)+\frac{15872 \zeta (3)}{3}
  \nn \\ &&  \hspace*{-62mm}
  \left.
   -\frac{5824 \pi ^2 \zeta
   (3)}{27}-\frac{4864 \log (2) \zeta (3)}{3}
  \right.  \nn \\ &&  \hspace*{-62mm}
 -\frac{21088}{27} \pi ^2 \log (2) \zeta(3)
   +\frac{448}{3} \log ^2(2) \zeta (3)
  \nn \\ &&  \hspace*{-62mm}
   +\frac{160}{9} \log ^3(2) \zeta (3)-\frac{16496 \zeta
   (3)^2}{9}-864 \zeta (5)
   \nn \\ &&  \hspace*{-62mm} \left.
   +2352 \log (2) \zeta (5)\right) \ep^3
+\ldots\,.
\nn
\eea
%%%%%%%%%%%%%%%%%%%%%%%%%%%%%%%%%%%%%%%%
In the above results the factor
$(i\pi^{d/2}e^{-\gm_E \ep})^3$ is implied on the right-hand side.
Furthermore, 
$s_6=\zeta(\{-5,-1\},\infty)+\zeta(6)
=\sum_{i=1}^{\infty}\frac{(-1)^{i}}{i^5}
\sum_{j=1}^{i}\frac{(-1)^{j}}{j}$, and
we show the $\ep$ expansion terms
up to the order which is needed for the static potential.
The results for $J_1, \ldots, J_{14}$ have been checked numerically with the
help of the program {\tt FIESTA}~\cite{FIESTA,Smirnov:2009pb}.

%- }}}
%- {{{ Result for a3:

\section{Result for $a_3$}

For completeness we repeat the result for the non-fermionic part of $a_3$,
$a_3^{(0)}$,
which has been obtained in Refs.~\cite{Smirnov:2009fh,Anzai:2009tm}
\begin{eqnarray*}
  a_3^{(0)} &=&
  502.24(1) \,\, C_A^3
  -136.39(12)\,\, \frac{d_F^{abcd}d_A^{abcd}}{N_A}
  \,,
%  \label{eq::a3}
\end{eqnarray*}
where $C_A=N_c$ and $d_F^{abcd}d_A^{abcd}/N_A = (N_c^3 + 6N_c)/48$ in the case
of $SU(N_c)$. Since for three master integrals the highest $\ep$ expansion
coefficient is only known numerically (see Ref.~\cite{Smirnov:2010zc})
no analytical result is available yet for $a_3^{(0)}$. Note, however, that the
achieved accuracy is sufficient for all foreseeable applications.

%- }}}
%- {{{ Acknowledgements:

\section*{Acknowledgements}

This work is supported by DFG through project SFB/TR~9
``Computational Particle Physics'' and RFBR, grant 08-02-01451.
V.S. appreciates the
financial support (for the participation in the workshop) of the Institute
%f\"ur Theoretische Teilchenphysik
of Theoretical Particle Physics
of KIT.

%- }}}

\end{document}